\documentclass[submission, Phys]{SciPost}

\hypersetup{
    colorlinks,
    linkcolor={red!50!black},
    citecolor={blue!50!black},
    urlcolor={blue!80!black}
}

\usepackage[bitstream-charter]{mathdesign}
\DeclareSymbolFont{usualmathcal}{OMS}{cmsy}{m}{n}
\DeclareSymbolFontAlphabet{\mathcal}{usualmathcal}

\begin{document}

\begin{center}{\Large \textbf{
Phase-Retrieval with Incomplete Autocorrelations Using Deep Convolutional Autoencoders\\
}}\end{center}

\begin{center}
Giovanni Pellegrini\textsuperscript{1,$\star$} and
Jacopo Bertolotti\textsuperscript{2,$\dagger$}
\end{center}

\begin{center}
{\bf 1} Dipartimento di Fisica, Universit\`{a} degli studi di Pavia, Via Bassi 6, Pavia 27100, Italy
\\
{\bf 2} Department of Physics and Astronomy, University of Exeter, Exeter, Devon EX4 4QL, UK
\\
${}^\star$ {\small \sf giovanni.pellegrini@unipv.it}, ${}^\dagger$ {\small \sf j.bertolotti@exeter.ac.uk}
\end{center}

\begin{center}
\today
\end{center}


\section*{Abstract}
{\bf
Phase-retrieval techniques aim to recover the original signal from just the modulus of its Fourier transform, which is usually much easier to measure than its phase, but the standard iterative techniques tend to fail if only part of the modulus information is available. We show that a neural network can be trained to perform phase retrieval using only incomplete information, and we discuss advantages and limitations of this approach.
}

\vspace{10pt}
\noindent\rule{\textwidth}{1pt}
\tableofcontents\thispagestyle{fancy}
\noindent\rule{\textwidth}{1pt}
\vspace{10pt}

\section{Introduction}
\label{sec:intro}
Scattering is one of the major limiting factors in imaging, as it scrambles the light forming the image we desire into a shapeless blob, and even moderate amounts of scattering in the optical path can easily deteriorate the achievable quality to an unacceptable level \cite{ping_introduction_2006,ishimaru_imaging_2012,ntziachristos_going_2010,bertolotti_non-invasive_2012}. A number of techniques have been developed to deal with it, each with its own set of advantages and disadvantages, and often working only in a very specific set of circumstances \cite{abramson_light--flight_1978,nasr_demonstration_2003,huang_optical_1991,strekalov_observation_1995,bennink_two-photon_2002,mosk_controlling_2012,katz_focusing_2011,kim_transmission_2015,dremeau_reference-less_2015}. One of them, based on the stellar speckle interferometry technique developed for ground-based astronomy \cite{dainty_stellar_1975}, exploits speckle correlations to measure the autocorrelation of the desired image, and then employs numerical techniques to invert the autocorrelation to yield the image itself \cite{fienup_phase_1982,bauschke_hybrid_2003,marchesini_alternating_2016}, a process known as ``phase retrieval'', which is usually achieved by employing one of the variations of the celebrated Gerchberg–Saxton algorithm \cite{gerchberg_practical_1972}. A common limiting factor of this approach is that it relies on the optical memory effect, i.e. the fact that the light from different point sources will produce the same (but shifted) speckle pattern when scattered. This is a very strong correlation, but it only works in a limited range, which decays exponentially with the scattering medium thickness \cite{freund_looking_1990}.

Among the approaches adopted to tackle the ``phase retrieval'' problem, deep-learning methods have gained increasing popularity in the recent years. If compared with traditional retrieval algorithms, they have shown much promise in a variety of domains ranging from non-line-of-sight imaging to ptychography \cite{metzler2020deep,cherukara_real-time_2018,cherukara_ai-enabled_2020,nguyenDeepLearningApproach2018, wang_phase_2020}, and they have proven to be particularly effective when dealing with reconstructions in presence of noise and scattering \cite{zhu2021imaging,zheng_non-line--sight_2021,tang_deepsci_2023}. Nevertheless, relatively little work has been done to directly address the field of view limitations imposed by the memory effect \cite{shi2022non}.

In this paper, we consider a reconstruction problem where we wish to recover the image of an hidden object $o$ from an estimate of the image autocorrelation $o \star o$, obtained from the analysis of the spatial correlations of the speckle image. Furthermore, we want to perform the image retrieval in  different conditions, where the autocorrelation information may be either fully available ($o \star o$) or, more interestingly, partially removed. In particular we will consider the case where only the autocorrelation within a disk $D(r_{m})$ of radius $r_{m}$ (the "mask radius") is available~\cite{metzler2020deep}. This scenario may arise in different experimental contexts, one of the most relevant being non-invasive imaging through strongly scattering media \cite{bertolotti_non-invasive_2012,katz_non-invasive_2014,metzler2020deep}.  A straightforward approach to tackle this problem would be to leverage the relationship $\mathcal{F}(o \star o) = |\mathcal{F}(o)|^{2}$, where $\mathcal{F}$ stands for the Fourier transform operation, and feed the obtained square amplitude to a standard phase retrieval algorithm. This technique is known to work reasonably well in presence of full information and in the absence of noise, but its performance progressively degrades as the input autocorrelation data become progressively non-ideal \cite{metzler2020deep}.
It is important to keep in mind that not all the information about the image is encoded in its autocorrelation. The absolute position is lost when performing an autocorrelation, so only the relative position between the different parts of the image can be reconstructed, and the autocorrelation of a real and positive function (like an image) is always centrosymmetric, so the reconstruction can not distinguish between an image and its centrosymmetric inverse.

In the following we investigate to which extent deep learning, and in particular Convolutional Neural Networks (CNN), can be employed to mitigate the averse effect of information removal on phase retrieval attempts. As a first step in this direction we verify how classic phase retrieval algorithms can deal with partially available autocorrelations $D(r_{m})(o \star o)$. Specifically, we adopt the Hybrid Input Output (HIO) algorithm as a reference approach \cite{fienup_reconstruction_1978,bauschke_phase_2002}, and we benchmark its performance on partial autocorrelations $D(r_{m})(o \star o)$ by applying a circular mask to erode the periphery of the full autocorrelation $o \star o$. Subsequently we devise a deep learning  workflow to tackle the same problem with tools traditionally provided by CNNs and autoencorder architectures \cite{metzler2020deep}.

\section{Methods}
\label{sec:methods}

\subsection{Traditional phase retrieval approach}
\label{sec:ph_re}
\begin{figure}[t]
\centering
    \includegraphics[width=0.45\textwidth]{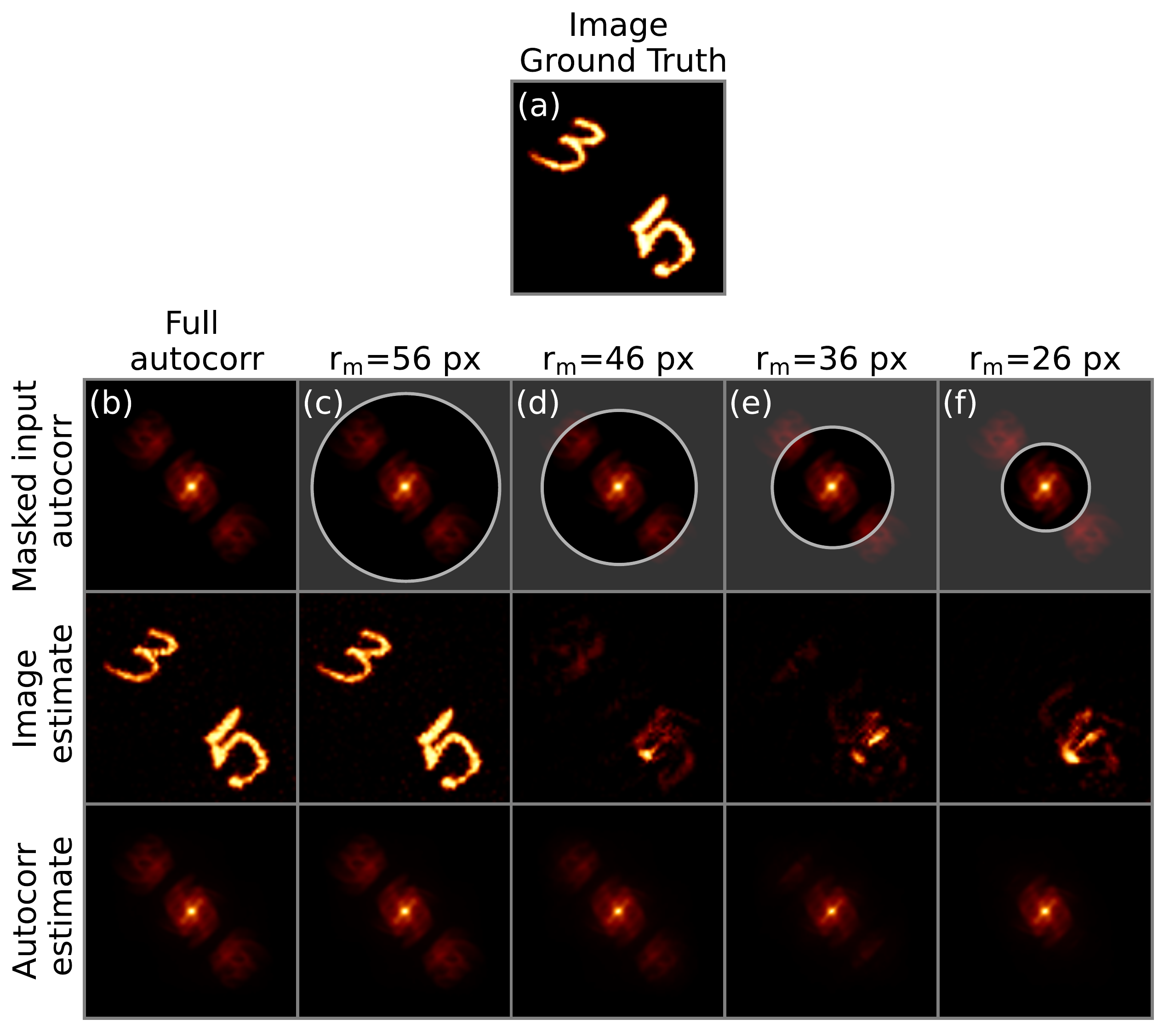}
    \caption{Autocorrelation reconstruction with HIO algorithm. (a) Image ground truth, (b) phase retrieval with full autocorrelation input, (c-f) phase retrieval after information removal with a $r_{m}=56$~px, $r_{m}=46$~px, $r_{m}=36$~px  and $r_{m}=26$~px circular mask.}
    \label{fig:hio}
\end{figure}

Traditional approaches, such as the Hybrid Input Output (HIO) algorithm \cite{fienup_reconstruction_1978,bauschke_phase_2002}, represent the state of the art in terms of classic image phase retrievals. It is instead unclear whether they can efficiently retrieve original images $o$ from partial autocorrelations $D(r_{m})(o \star o)$. We test this assertion by implementing the HIO algorithm and, starting from a random guess, running it for 400 iterations and 20 trial runs for each phase retrieval attempt, both for full and partial autocorrelation inputs. 
As a sample input, we choose a simple 128x128 pixels image representing two handwritten digits, and the corresponding autocorrelation.
The results of these retrieval attempts are reported in Figure~\ref{fig:hio}. In this instance, we obtain a successful phase retrieval only when a complete autocorrelation is fed to the HIO algorithm (Figure~\ref{fig:hio}b). If instead we apply a circular erosion mask to the same input autocorrelation, the reconstruction succeeds only if the mask application does not result in any information removal (Figure~\ref{fig:hio}c). On the other hand, as soon as the masking leads to the smallest removal of information, the quality of the image retrieval degrades dramatically (Figure~\ref{fig:hio}d-f). Eventually, a large enough information erosion degrades the phase retrieval process up to a point where the original image is no longer recognizable in the retrieved solution. This is because traditional approaches such as HIO can only target the available portion of the autocorrelation, which in turn corresponds to a source image completely different from the original one.

\subsection{Deep learning: dataset}
\label{sec:dl_data}
\begin{figure}[t]
\centering
    \includegraphics[width=0.45\textwidth]{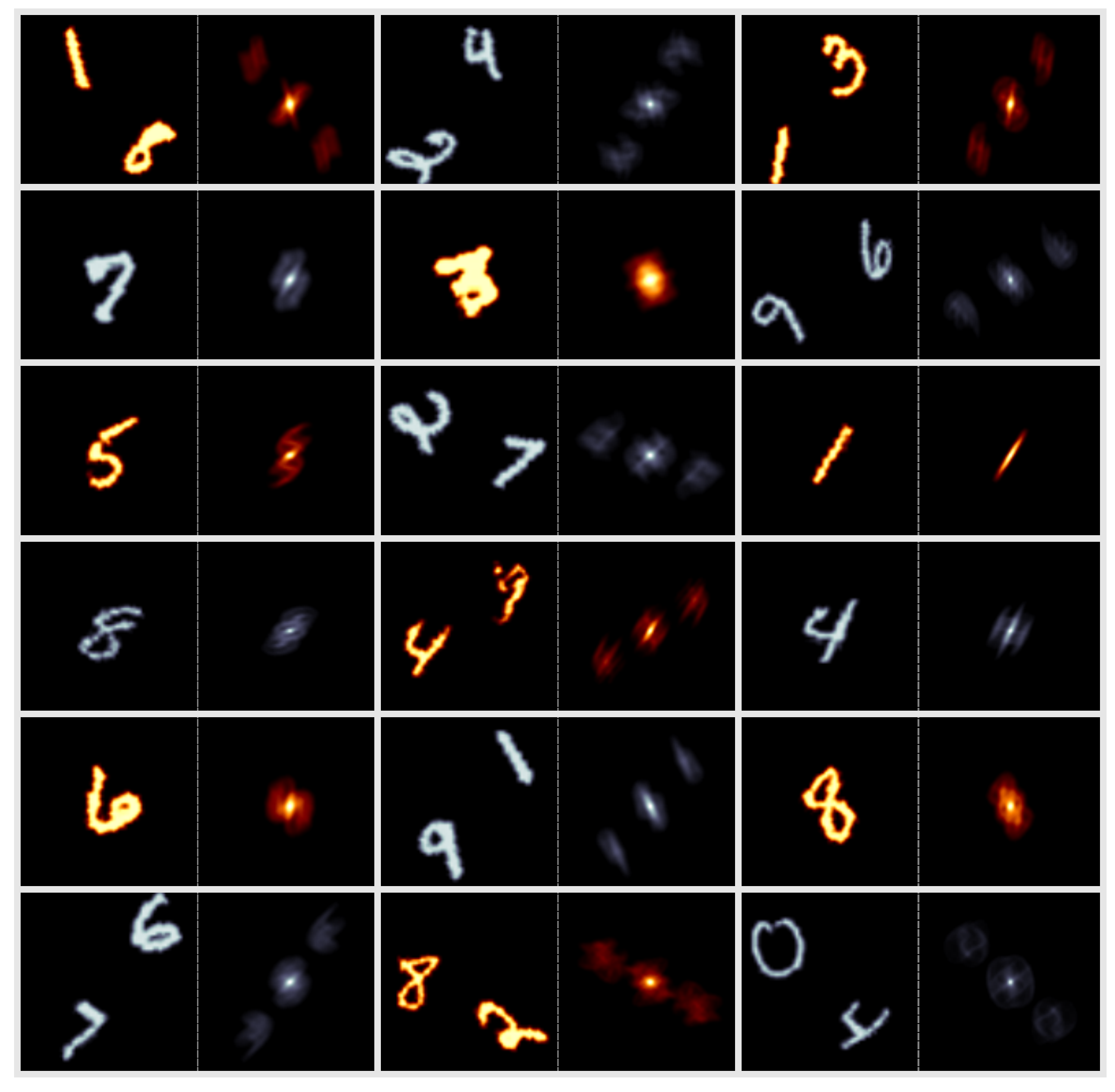}
    \caption{Samples drawn from the synthetic dataset. Each sample of the dataset is formed by and image-autocorrelation pair. We alternate two different color scales to better distinguish the dataset entries. Each ground truth image contains a single digit or a pair of digits drawn from the MNIST hand written digit database.}
    \label{fig:dataset}
\end{figure}

In the following, we devise an efficient deep learning workflow for the solution of the phase retrieval problem. The first step is the construction of an appropriate synthetic dataset. We choose the MNIST handwritten digits database as a starting point to generate suitable samples to address the partial autocorrelation reconstruction problem \cite{lecun1998gradient}. Overall, each entry of the synthetic dataset consists of an image, composed of one or more digits ($o$), and the corresponding full autocorrelation ($o \star o$); a few representative examples of the constructed dataset are reported in  Figure~\ref{fig:dataset}. In practice, half of the dataset are pairs of MNIST digits, randomly combined after a rotation in the $[-\pi/4,\pi/4]$ interval (so that centrosymmetric inversions in the reconstructions are still clearly visible at a glance), and matched with the corresponding autocorrelation. The remaining half consists instead of single digit image-autocorrelation pairs, where again each digit is randomly rotated in the same angular interval. We generate a training dataset of 200000 samples and a corresponding validation dataset of 10000 samples, and note that digits drawn from the MNIST training set will exclusively populate our synthetic training set, and likewise for the generation of the validation set. We finally underline that such a dataset has no ambition to represent the generality of all situations encountered in phase retrieval problems, but we believe that it can provide useful indications regarding the potential of deep learning approaches for the reconstruction of images starting from partially available information.

\subsection{Deep learning: architecture, loss and training procedure}
\label{sec:dl_arch}
\begin{figure*}[t]
\centering
    \includegraphics[width=0.7\textwidth]{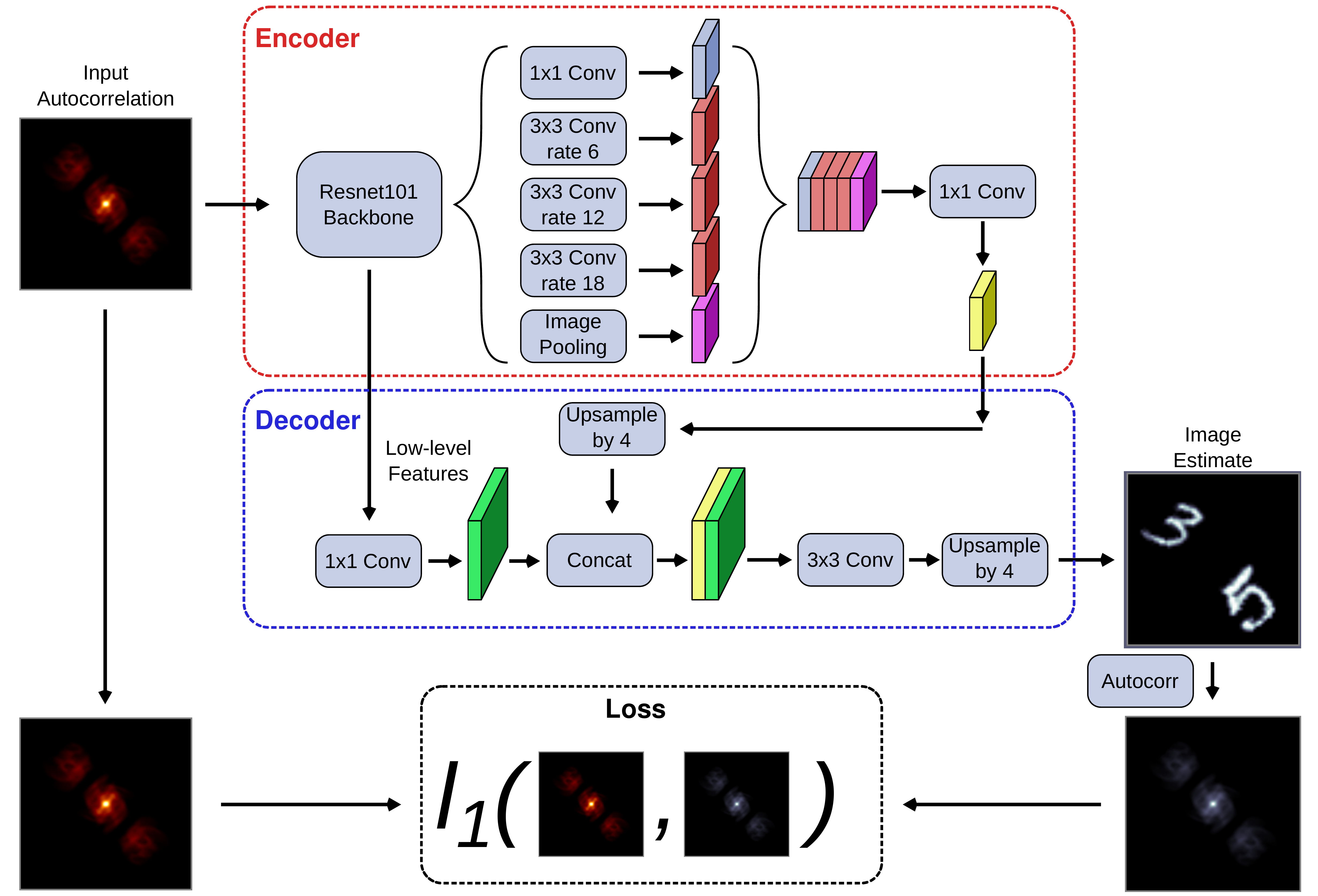}
    \caption{DeepLabV3+ model schematics. Encoder: we employ a ResNet101 backbone, sample and stack the extracted features through atrous spatial pyramid pooling (ASPP) and compress them with a 1x1 convolution. Decoder: we concatenate the compressed ResNet101 features with the ASPP output and apply a 3x3 convolution followed by an upsampling to match the desired output size. Loss: We compute the autocorrelation of the reconstructed image and compute an $l_{1}$ loss with the input ground truth autocorrelation.}
    \label{fig:model}
\end{figure*}

We adopt the DeepLabV3+ model with a ResNet101 backbone for the phase retrieval task \cite{chen2018encoder}. The DeeplabV3+ network is an encoder-decoder CNN architecture, originally employed to tackle semantic segmentation tasks, which can be adapted to several other applications (Figure~\ref{fig:model}). If compared to other more traditional CNN models, DeepLabV3+ has the advantage to probe convolutional features at multiple scales and to provide a denser and less compressed feature extraction at the encoder level, while keeping the decoder structure extremely simple \cite{chen2018encoder}.

In our workflow, the model is simply fed either a full or a partial autocorrelation, and outputs the corresponding ground truth image. We guide the training computing an ordinary $l_{1}$ loss between the autocorrelation ground truth and the reconstructed autocorrelation, to avoid the memorization of the digit locations in the training data (Figure~\ref{fig:model}) \cite{metzler2020deep}.

We train our network using Stochastic Gradient Descent with momentum (SGD), a variable learning rate and a batch size of 128. We implement the solution in Pytorch and Pytorch-Lightning and train the network for about 5000 epochs using 8 Nvidia V100 GPUs over the time of one week \cite{NEURIPS2019_9015,Falcon_PyTorch_Lightning_2019}.

\begin{figure}[t]
\centering
    \includegraphics[width=0.45\textwidth]{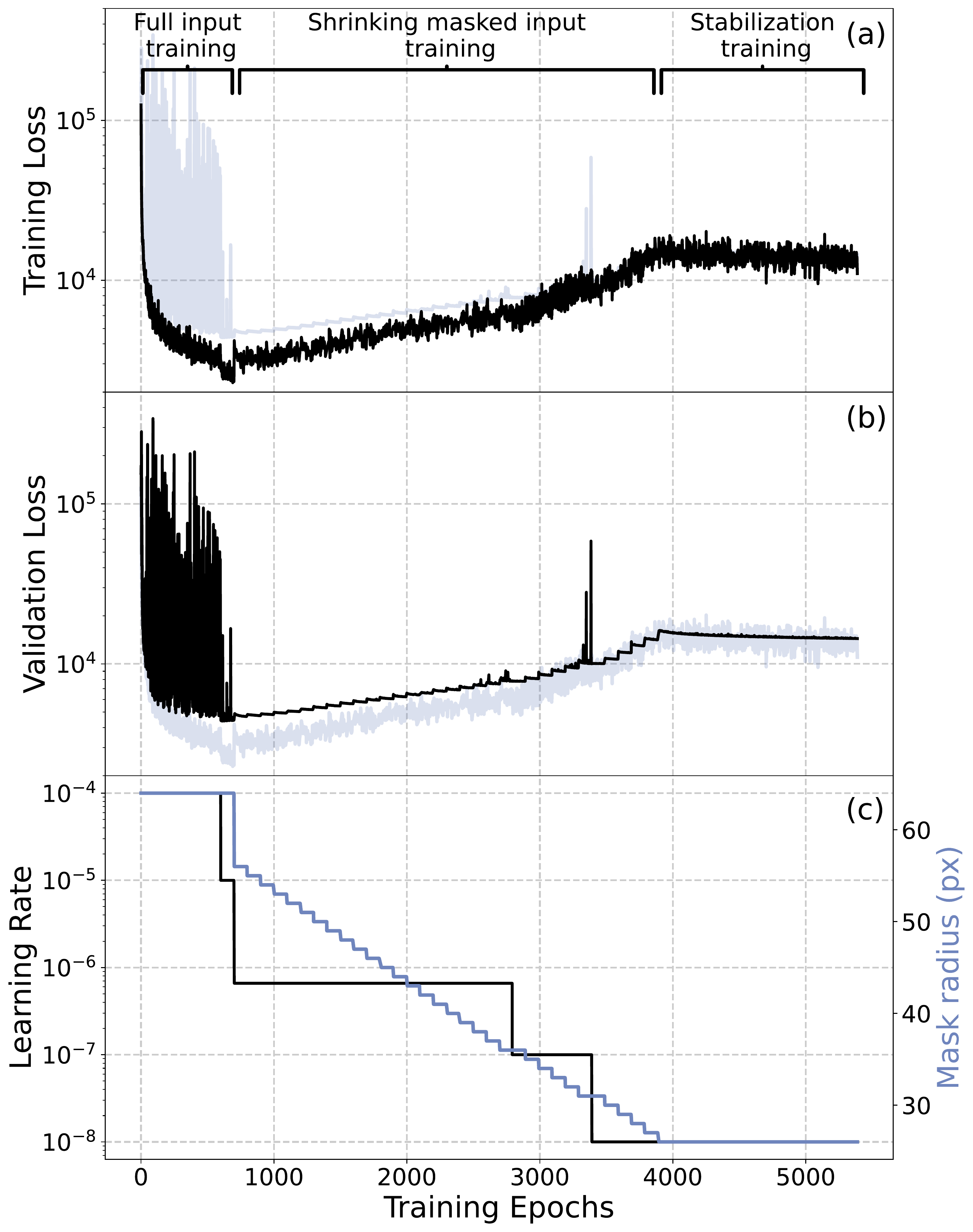}
    \caption{Training curves. (a) Training loss (black line) and validation loss (light gray line) added for comparison. (b) Validation loss (black line) and training loss (light gray line) added for comparison. (c) Learning rate (black line) and mask radius (blue line) schedules.}
    \label{fig:training}
\end{figure}

The overall training procedure is divided in three main stages, as shown in Figure~\ref{fig:training}. During the first training stage, lasting 700 epochs, we train the network to reconstruct the ground truth image from the full autocorrelation with a learning rate equal to $l_{r}=10^{-4}$, and scale it down to $l_{r}=10^{-5}$ for the last 100 epochs to reduce the noise in the validation loss.

We then transition to the second, 3200 epochs long, training phase. At this stage, we apply a circular erosion mask to the input autocorrelation, starting with a mask radius of $r_{m}=56$~px and terminating with a radius of $r_{m}=26$~px, shrinking $r_{m}$ of one pixel every 100 epochs. During this training stage it is necessary to reduce the learning rate from an initial value of $l_{r}=10^{-6}$ to a final value of $l_{r}=10^{-8}$ in order to keep the validation loss landscape sufficiently stable and reasonably free of noise. It is clear that, as shown in Figure~\ref{fig:training}, each pixel erosion causes an uptick in the training and validation loss, followed by a partial recovery during the remaining training epochs. This means that the information removal negatively impacts the reconstruction performance, but the adverse effect can be partially recovered during the training process.
Finally, in the third and last stage, we run the training for an additional 1500 epochs with a constant learning rate of $l_{r}=10^{-8}$.

We extract several network checkpoints during the training process to monitor the network phase retrieval performance at different stages of information erosion. We archive the first checkpoint at the end of the full autocorrelation training, and save three more for an autocorrelation mask size of 56, 46 and 36 pixels. Finally, we archive the checkpoints at the end of the training, corresponding to a mask size of 26 pixel. All of our training and inference code is made available in a public repository \cite{giovanni_pellegrini_2023_7842075}.

\section{Results and Discussion}
\label{sec:res_dis}
It is now interesting to explore the phase retrieval performance of the deep learning approach at different stages of information erosion, and compare it with that of a traditional algorithm.

\begin{figure}[t]
\centering
    \includegraphics[width=0.45\textwidth]{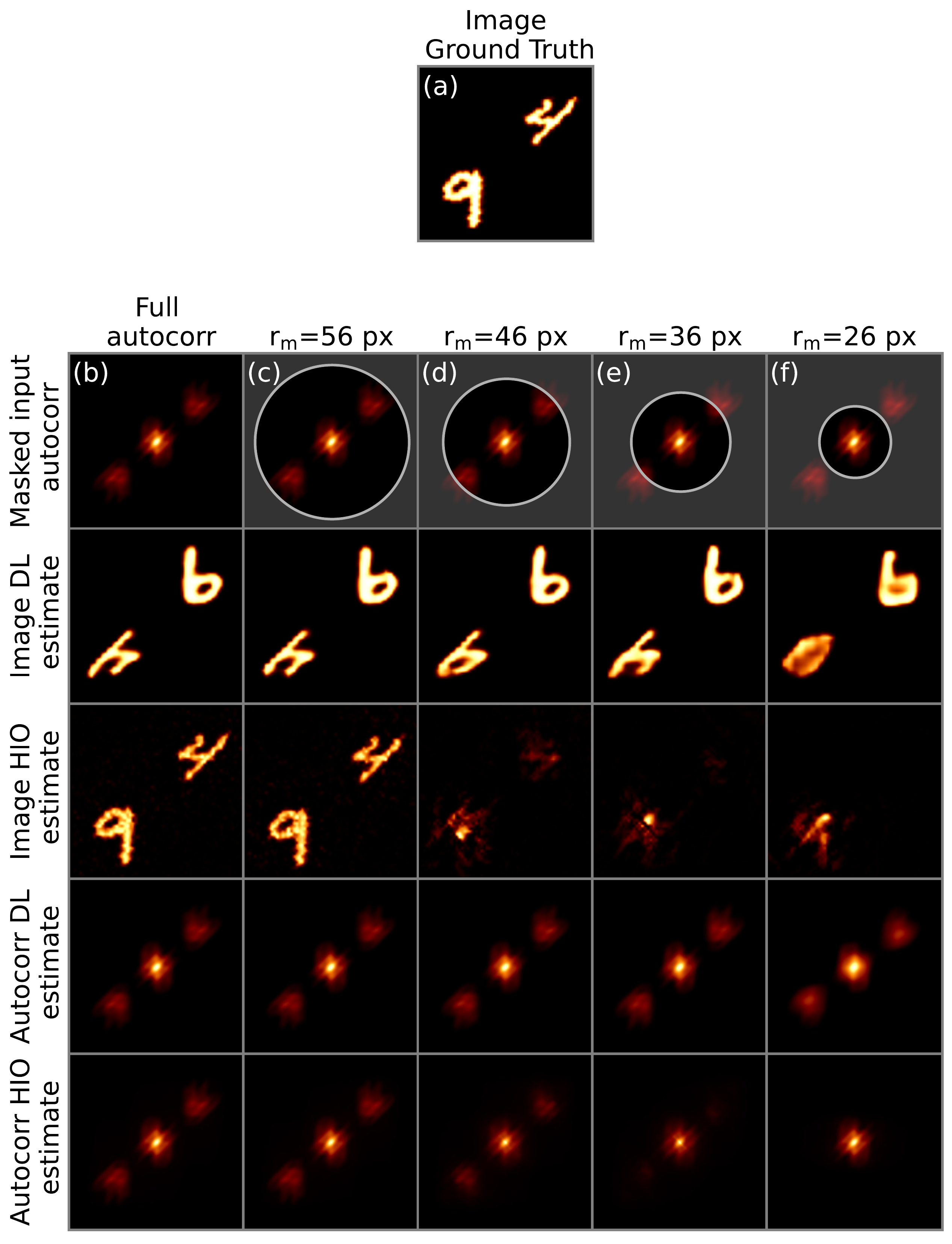}
    \caption{Performance comparison between HIO and deep learning algorithms. (a) Image ground truth, (b) phase retrieval with full autocorrelation input, (c-f) phase retrieval after information removal with a $r_{m}=56$~px, $r_{m}=46$~px, $r_{m}=36$~px  and $r_{m}=26$~px circular mask.}
    \label{fig:DeepVsHio}
\end{figure}

Figure~\ref{fig:DeepVsHio} examines the phase reconstruction performance of the deep learning and the HIO approaches for a random sample drawn from our validation dataset. When dealing with a full autocorrelation input, Figure~\ref{fig:DeepVsHio}b shows that both techniques display a similar performance, even though we notice that, also in this trivial case, the CNN output is less noisy and closer to the ground truth image. In this instance, we underline that the image reconstructed by neural network, while correct, is flipped by a point symmetry if compared to the original one, since the loss functions targets the difference between autocorrelations and is invariant under center point transformations.

As we progress with the information erosion, and shrink the mask radius below the $r_{m}=56$~px limit, the difference in the quality of the phase retrieval between the deep learning and the traditional approach becomes apparent. If, as an example, we explore the phase retrieval attempt for $r_{m}=46$~px (Figure~\ref{fig:DeepVsHio}d), it is indeed clear that the image reconstruction with the deep learning approach has undergone little to no degradation, while the traditional solution relying on the HIO algorithm has completely broken down.

The same is true for the $r_{m}=36$~px mask application (Figure~\ref{fig:DeepVsHio}e), with the deep learning reconstruction virtually identical to the ground truth, and the HIO algorithm unable to produce a meaningful output.

The situation is instead qualitatively different for the largest information erosion obtained applying the $r_{m}=26$~px mask. Figure \ref{fig:DeepVsHio}f highlights a situation where, while the traditional phase retrieval is still unable to provide a proper reconstruction, the deep learning approach also struggles to perform accurately. In this instance the CNN correctly locates the position of the digits inside the image, but cannot accurately reconstruct the details of each one, resulting in an overall blurred reconstruction. This lack of accuracy is also reflected in the computed autocorrelation, which is correctly reconstructed in broad strokes, but lacks the fine details of the target ground truth. The fact that the degradation in the output coincides with the next to complete loss of the side lobes in the autocorrelation, suggests that, contrary to more traditional algorithms, a deep learning approach can still produce reliable results as long as some information from the autocorrelation side lobes (which encode the spatial relation between different objects in the scene) is retained.\\

As a final investigation, we wish to study the behavior of the deep learning phase retrieval approach for a variety of different inputs, including single digit images and out of distribution samples, such as images containing never observed letters from the alphabet.

\begin{figure}[t]
\centering
    \includegraphics[width=0.45\textwidth]{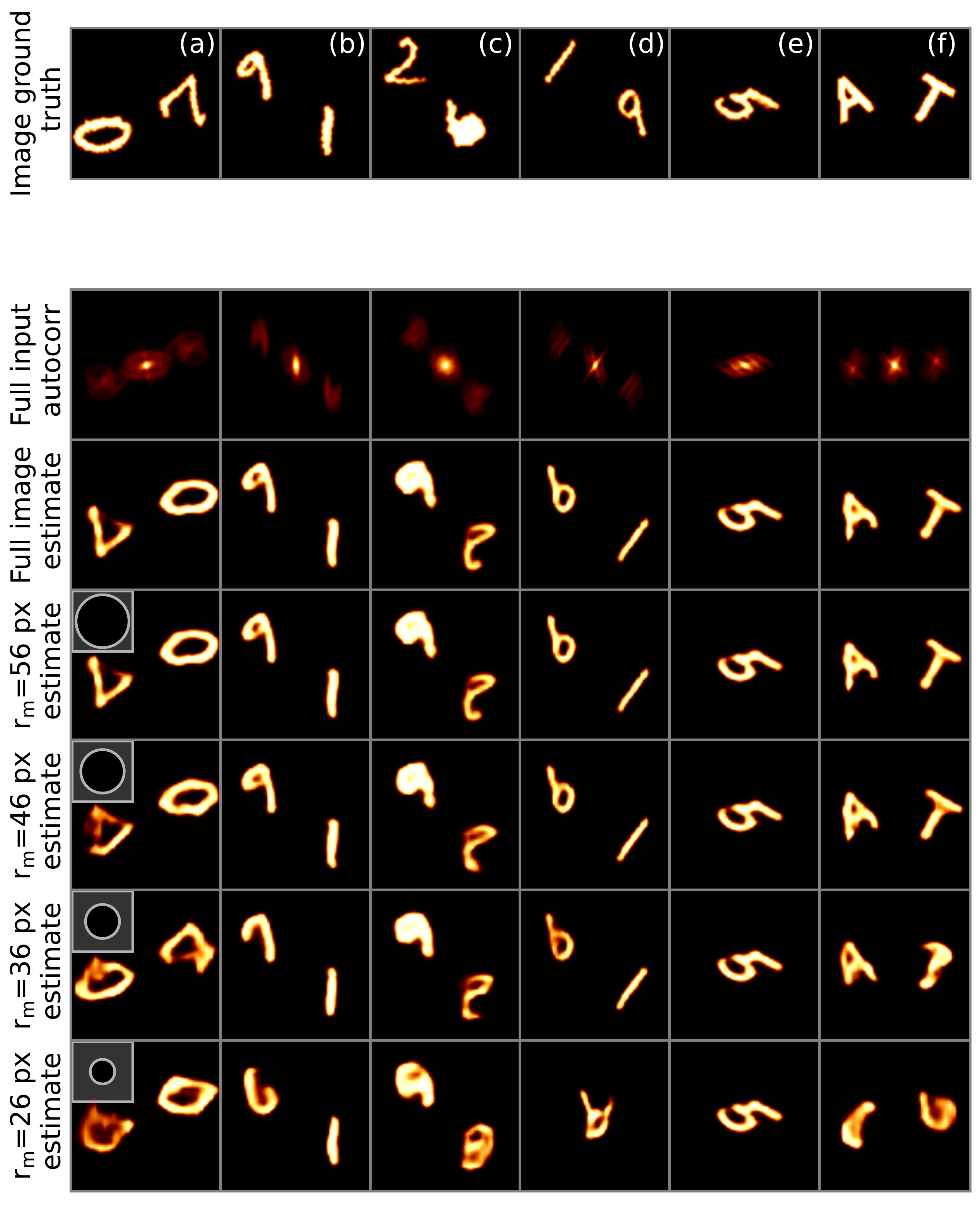}
    \caption{Examples of deep learning reconstruction for different degrees of masking: (a,c) digit reconstruction degradation for the $r_{m}=26$~px mask, (b) single digit flip for the $r_{m}=26$~px mask, (d) collapse of the image reconstruction for the $r_{m}=26$~px mask, (e) successful reconstruction of a single digit image and (f) successful reconstruction of an out of distribution sample containing letters of the alphabet.}
    \label{fig:DeepSampling}
\end{figure}
Figure~\ref{fig:DeepSampling} displays a selection of these phase retrieval attempts. The first four examples reported in Figure~\ref{fig:DeepSampling}a-d are drawn from the validation set, and roughly follow the trends already observed in Figure~\ref{fig:DeepVsHio}. The last two cases of Figure~\ref{fig:DeepSampling}e-f are qualitatively different single digit and out of distribution inputs.
As before, the deep learning approach provides a satisfactory performance up to the $r_{m}= 46$~px and $r_{m}=36$~px erosion stage. For more extremes crops of the autocorrelation, some information is irremediably lost, and the reconstruction suffers from that. This might show up as a deviation from the shape of the ground truth (e.g. Figure~\ref{fig:DeepSampling}a, c, or f), lack of information about the relative distance between the different parts of the image and thus a collapse of two digits into a single shape (e.g. Figure~\ref{fig:DeepSampling}d), or sometimes more subtle mistakes like the flip of one digit, but not the whole figure, in Figure~\ref{fig:DeepSampling}b.
This analysis is consistent with recent results reported in the literature, where a CNN can successfully reconstruct an image ground truth composed of two digits from the center lobe of the autocorrelation, but only when given constraints about the number and shape the of included digits \cite{shi2022non}.
We finally remark that the network successfully reconstructs single digits autocorrelations without hallucinating any spurious side lobes (Figure~\ref{fig:DeepSampling}e) and that even out of distribution samples are handled correctly (Figure~\ref{fig:DeepSampling}f).

\section{Conclusion}
\label{sec:conc}
The presented results indicate that, whereas traditional phase retrieval algorithms struggle to perform efficiently if presented partial information, CNNs can overcome the detrimental effects of the information removal, and largely exceed the performance achievable with conventional procedures. The deep learning approach easily performs phase retrievals in regimes where classical methods fail to deliver any meaningful result. In practice, CNNs can reconstruct multiple digits images from partial autocorrelations in which the information about the number and relative position of the digits has been almost completely removed.
At the same time it is important to always remember that the training dataset biases the output, so such a system can only ever be reliable on cases close enough to the one it was trained on, and can easily hallucinate superficially plausible reconstructions very different from the ground truth~\cite{Fitzpatrick2021}.

\bibliography{SciPost.bib}

\nolinenumbers

\end{document}